\documentclass[useAMS,usenatbib]{mn2e}
\usepackage{times}
\usepackage{amssymb}
\usepackage{graphicx}
\begin{document}

\title[Thermal generation of the magnetic field 
       in massive stars]
      {Thermal generation of the magnetic field 
       in the surface layers of massive stars} 
\author[V. Urpin]
  {V.~Urpin$^{1,2)}$ \\
      $^{1}$ INAF, Osservatorio Astrofisico di Catania,
      Via S.Sofia 78, 95123 Catania, Italy \\
      $^{2)}$ A.F.Ioffe Institute of Physics and Technology,
      194021 St. Petersburg, Russia}
\date{today}

\pagerange{\pageref{firstpage}--\pageref{lastpage}} \pubyear{2002}

\def\LaTeX{L\kern-.36em\raise.3ex\hbox{a}\kern-.15em
    T\kern-.1667em\lower.7ex\hbox{E}\kern-.125emX}

\newtheorem{theorem}{Theorem}[section]

\label{firstpage}

\maketitle

\begin{abstract}
A new magnetic field-generation mechanism based on the Nernst effect
is considered in hot massive stars. This mechanism can operate in the 
upper atmospheres of O and B stars where departures from the LTE form 
a region with the inverse temperature gradient. 
\end{abstract}


\section{Introduction}

The investigations of stellar magnetism and its influence on the 
stellar surfaces is a quickly maturing research area. Recently, 
a number of studies have been carried out on a detection of the 
magnetic field in massive B- and O-type stars (see, e.g., Sch\"oller 
et al. 2016, Hubrig et al. 2016). While in a number of B stars, 
the magnetic field was detected already several decades ago, the 
existence of magnetic fields in O stars has been proven only recently. 
The direct field measurements are difficult in such stars since only 
a few spectral lines are available for this. Besides, the average 
magnetic field of O stars turns out to be $\sim 8-9$ times weaker 
than the field of B and A stars (Kholtygin et al. 2015). 

The origin of magnetic fields in massive stars is still poorly
understood. Upper main-sequence stars with $M > 2 M_{\odot}$ have 
no convective envelopes and it seems that dynamo should not be 
efficient in OB stars. Therefore, some authors argue that the magnetic 
field of these stars is fossil or generated by some exotic mechanisms 
(like, e.g., interaction in stellar mergers or in the course of a mass 
transfer). However, the role of dynamo can be underestimated. For 
instance, Cantiello \& Braithwaite (2011) considered the 
subsurface mechanism, assuming that dynamo generates the magnetic 
field in the Fe convection zone. They found that the field generated 
in this zone could emerge at the surface via magnetic buoyancy and be 
widespread. Also, it has been shown that a significant fraction of OB 
stars may suffer merger or mass-transfer events in their evolution 
and the field can be generated through such processes (e.g., Sana et 
al. 2012, Wickramasinghe et al. 2015). 

Dynamo, however, is not the only mechanism that amplifies the 
magnetic field in astrophysical bodies. Sometimes, the electric 
currents supporting the magnetic field can be related to other 
mechanisms. For instance, the Biermann battery is an example 
of such non-dynamo mechanism (see, e.g., Biermann 1950, Kemp 
1982, Mestel \& Moss 1983) that operates in plasma if gradients 
of the pressure and number density of electrons are non-parallel. 
Several astrophysical applications of this process have been 
discussed that range from stellar magnetic fields to seed magnetic 
fields on galactic scales. For example, this process can generate 
a seed magnetic field around first stars (see, e.g., Ando et al. 
2010, Doi \& Susa 2011). Note that a similar mechanism (often 
called the cross-helicity effect) operates in turbulent plasma 
(Yoshizava 1990, Brandenburg \& Urpin 1998). In this case, the role 
of the pressure gradient is played by the gradient of turbulent 
stresses.

In this paper, we consider the mechanism of a magnetic field 
generation that operates in plasma with a non-uniform temperature.  
This mechanism is caused by the so-called thermomagnetic 
instability and involves a range of processes operating in hot 
plasmas. The thermomagnetic processes transform 
a fraction of the thermal flux into the energy of magnetic fields. 
Such possibility first was considered with regards to experiments 
in the laser plasma (see, e.g., Tidman \& Shanny 1974, Bol'shov et 
al. 1974, Dolginov \& Urpin 1979). In this scenario, a feedback acts 
between the Nernst effect and the Righi-Leduc heat flow and is 
determined entirely by transport processes and neither hydrodynamic 
motions nor density gradients are required. In the laser plasma, 
the thermomagnetic processes lead to the instability that generates 
strong magnetic fields ($\sim 10^6 - 10^7$ G) on a short timescale
(see, e.g., Haines 1981, Andrushchenko \& Pavlenko 2004). The impact
of both density gradients and hydrodynamic motions on the thermomagnetic 
instability is considered by Bissell et al. 2010, 2012. Hydrodynamic
motion is found to produce a limited effect on the growth rate, whereas
the influence of density gradients can be more essential because they
produce an additional source term. In un-magnetised plasmas, it is 
widely believed that this instability may be driven by two mechanisms 
which are determined by i) non-parallel gradients of the temperature 
and electron number density and ii) the Nernst advection that can lead 
to exponential compression of the magnetic perturbations (see Bissell  
2015 for more detail). Note that the Nernst advection can contribute
to the field generation if temperature and density gradients are 
parallel and if this condition fails (Brownell 1979, Hirao \& 
Ogasawara 1981).

Under astrophysical conditions, this instability has been considered in 
the context of the early evolution of neutron stars where it generates 
strong magnetic fields in the surface layers (see Urpin et al. 
1986). The evolution of thermally generated magnetic fields can 
account for a number of qualitative features observed in 
pulsars (such as fast decay at the early evolution and slow 
decay during the later evolution; see Urpin \& van Riper 1993, 
Urpin et al. 1994). Note that thermomagnetic processes can occur 
in non-magnetized as well as magnetized plasma, and motions 
caused by these processes influence transport properties of plasma.
The mechanism for such instability in magnetized plasmas is well
studied. Generally, feedback between the Nernst effect and
the Righi-Leduc heat flow can be caused by transport processes and 
neither hydrodynamic motions nor density gradients are required. 
The instability yields propagating magnetothermal waves whose direction 
depends on the magnitude of the Hall parameter. 
For instance, these processes are important in hot accretion discs 
(Montani et al. 2013, Urpin \& R\"udiger 2005, Franko et al. 2014) 
where the thermomagnetic phenomena are accompanied by 
the magnetorotational ones and contribute to the angular 
momentum transport. 

In this paper, we consider the thermomagnetic processes that operate
in the surface layers of massive stars. Plasma in the upper 
atmosphere of OB stars is hot and its density is relatively low. 
Therefore, thermomagnetic phenomena can lead to an efficient 
generation of the magnetic field. We argue that characteristics of 
the magnetic field generated by this mechanism in massive stars 
are in qualitative agreement with observational data.

\section{Basic equations}

Consider the thermomagnetic instability in the upper atmosphere
by making use of a plane-parallel geometry with $(x, y, z)$ being 
the Cartesian coordinates. We assume that within the layer between 
$z=0$ (bottom) and $z=a$ (top), the temperature gradient is directed 
in a positive or negative $z$-direction. 
The generalized Ohm's law in fully ionised plasma reads (see, 
e.g., Braginskii 1965)
\begin{equation}
\vec{E} = - \frac{\vec{v}}{c} \times \vec{B} - \frac{\vec{B} \times 
\vec{j}}{e n_e} - \frac{\nabla p_e}{e n_e} + \frac{\hat{\alpha} \cdot 
\vec{j}}{(e n_e)^2} - \frac{\hat{\beta} \cdot \nabla T}{e n_e},
\end{equation} 
where $\vec{E}$, $\vec{B}$, and $\vec{j}$ are the electric and 
magnetic fields and the electric current,  $\vec{v}$ is the velocity, 
$n_e$ and $p_e$ are the number density and pressure of electrons, 
respectively; $T$ is the temperature; $e$ is the electron charge, 
$\hat{\alpha}$ and $\hat{\beta}$ are tensors that characterize 
the rate of dissipation of electric currents and the efficiency 
of thermomagnetic phenomena. The tensor productions in Eq.~(1) 
read 
\begin{eqnarray}
\hat{\alpha} \cdot \vec{j} = \alpha_{\parallel} j_{\parallel} + 
\alpha_{\perp}
j_{\perp} - \alpha_{\wedge} \vec{b} \times \vec{j},
\nonumber \\
\hat{\beta} \cdot \nabla T = \beta_{\parallel} \nabla_{\parallel} T + 
\beta_{\perp} \nabla_{\perp} T + \beta_{\wedge} \vec{b} \times 
\nabla T,
\end{eqnarray}  
where $\vec{b} = \vec{B}/B$; the subscripts $\parallel$, $\perp$, 
and $\wedge$ mark the components parallel and perpendicular to 
the magnetic field and the so-called Hall component. The components 
of $\hat{\alpha}$ and $\hat{\beta}$ depend on the parameters 
of plasma and its chemical composition. For one component fully 
ionized plasma, these coefficients have been calculated by 
Braginskii (1965). The kinetic coefficients in 
a multicomponent plasma have been considered by Urpin (1981).   

Combining the Faraday's law  $\partial \vec{B}/ \partial t = - c 
\nabla \times \vec{E}$ with Eq.~(1), we obtain the induction 
equation
\begin{eqnarray}
\frac{\partial \vec{B}}{\partial t} - \nabla \! \times \! (\vec{v} \! 
\times \! \vec{B})- \frac{c}{e} \nabla \! \times \! \left( 
\frac{\vec{B} \! \times \! \vec{j}}{n_e} \right) -
\frac{c}{e} \nabla \! \times \! \left( \frac{\nabla p_e}{n_e} \right)
\nonumber \\
+ \frac{c}{e^2} \nabla \! \times \! \left( \frac{1}{n_e^2} 
\hat{\alpha} \cdot \vec{j} \right) - \frac{c}{e} \nabla \! 
\times \! \left( \frac{1}{n_e} \hat{\beta} \cdot \nabla T 
\right) =0. 
\end{eqnarray}
Consider a particular type of waves governed by this equation.
We use the standard linear approach that is valid for waves with 
small amplitude and we represent all quantities 
as a sum of the unperturbed quantity and a small disturbance that 
will be marked by a subscript 1. The disturbances are governed 
by linearized MHD equations while they are small compared 
to the unperturbed quantities. We assume that unperturbed velocity 
and magnetic field are vanishing. 
Linearizing tensors $\hat{\alpha}$ and $\hat{\beta}$, we take 
into account that all terms $\propto B^2$ should be neglected 
because there is no magnetic field in the basic state. Then, 
we have for fully ionized hydrogen plasma 
\begin{eqnarray}
\alpha_{\parallel} \approx \alpha_{\perp} = 0.51 \frac{m_e n_e}{\tau_e} 
\;,\;\;\; \beta_{\parallel} \approx \beta_{\perp} = 0.71 n_e k_B \;, 
\nonumber \\
\beta_{\wedge} = 0.81 \frac{e n_e k_B \tau_e}{m_e c} B_1,  
\end{eqnarray}
where $k_B$ is the Boltzmann constant and $\tau_e$ is the 
electron relaxation time (Braginskii 1965). In fully ionized 
hydrogen plasma, the relaxation time of electrons is given by 
$\tau_e = 3 \sqrt{m_e} (k_B T)^{3/2}/ 4 \sqrt{2 \pi} e^4 n_e \Lambda$ 
where $\Lambda$ is the Coulomb logarithm (see, e.g., Spitzer 1998). 
The coefficient $\alpha_{\wedge}$ is proportional to $B_1$ but it should 
be multiplyed in Eq.(3) by the electrc current that is $ \vec{j}_1
= (c/4 \pi) \nabla \times \vec{B}_1$ and, hence, this term is 
non-linear in a small disturbance $B_1$ and must be neglected. 
Therefore, the linearized induction equation reads
\begin{eqnarray}
\frac{\partial \vec{B}_1}{\partial t} = 
- \frac{c^2}{4 \pi} \nabla \times \left( \frac{1}{\sigma}
\nabla \times \vec{B}_1 \right) +
\nonumber \\ 
\frac{c}{e} \nabla \times \left( \frac{\nabla p_{e1}}{n_e} -
\frac{n_{e1} \nabla p_e}{n_e^2} \right) -
0.81 \frac{k_B}{m_e} \nabla \times (\tau_e \nabla T \times \vec{B}_1), 
\end{eqnarray}   
where $\sigma = e^2 n_e \tau_e / 0.51 m_e$ is the conductivity
along the magnetic field. The first term on the r.h.s. of Eq.(5)
describes the standart Ohmic dissipation that always operates in
plasma, the second term describes magnetic field generation by the 
Biermann battery ($\propto \nabla T \times \nabla n_e$), and 
the third term corresponds to Nernst advection of the magnetic field 
lines. 

Eq.~(5) should be complemented by the heat balance, momentum, 
and continuity equations. The heat equation reads
\begin{equation}
\rho c_p \frac{d T}{d t} - \frac{d p}{dt} = - \nabla \cdot \vec{q_e}
- \nabla \cdot \vec{Q} + G - \Lambda, 
\end{equation}
where $\rho$ and $p$ are the density and total pressure, respectively; 
$c_p$ is the specific heat for $p$=const; $G$ and $\Lambda$ are 
the heating and cooling rates; $\vec{q_e}= - \hat{\kappa_e} \cdot 
\nabla T$ is the heat flux transported by electrons with $\hat{\kappa_e}$ 
being the tensor of electron thermal conductivity; $\vec{Q}$ 
describes the radiative heat flux; $d/dt = \partial / \partial t 
+ (\vec{v} \cdot \nabla)$. The tensor of 
electron thermal conductivity is given by the standard expression 
(see, e.g., Braginski 1965). The radiative heat flux, $\vec{Q}$, 
has a simple form in optically thick layers where $\vec{Q} = -\kappa_r 
\nabla T$ with $\kappa_r$ being the radiation thermal conductivity.  
Note that this simple expression does not describe the radiative flux 
in a region with the optical depth $\leq 1$. 

In the optically thin region, the radiative flux has a more complicated
shape and should be considered by making use of the radiative 
transfer equation. 
The expression for $\vec{Q}$ in Eq.~(6) is determined obviously 
by the thermodynamic parameters of plasma. Various approximate 
approaches have been suggested to describe this quantity in 
optically thin regions. 
For our purposes, we can choose the approach used by Wang 
(1966) and Ojha (1987). Following Wang (1966), the quantity 
$\vec{Q}$ can be represented as a power low function of 
the density and temperature,
\begin{equation}
\nabla \cdot \vec{Q} \approx  a c W_0 \rho^{1 + \zeta} 
T^{4 + \iota},          
\end{equation}
with $W_0$, $\zeta$, and $\iota$ being constants determined 
by fitting with opacity data. Note that the results of our 
study are not sensitive to a particular shape of the dependence 
of $\vec{Q}$ on $\rho$ and $T$. The only important point is 
that $\vec{Q}$ is weakly dependent on $B$ but this is certainly 
the case in stellar atmospheres where the magnetic field is of 
the order of $10^2-10^3$ G because radiative opacities are not 
sensitive to $B$ for such fields. The absorption coefficient 
becomes to be dependent on $B$ only in a much stronger field 
$\sim 10^8-10^9$ G. If $\vec{Q}$ depends weakly on $B$ then our 
conclusions are the same for any dependence of $\vec{Q}$ on $\rho$ 
and $T$.

The thermomagnetic effects are determined by the temperature
gradient rather than the electron heat-flux. In stellar atmospheres, 
the heat transport by electrons is much less efficient than the 
radiative transport and $Q \gg q_e$. For instance, it can be  
estimated that $Q$ is more than 15 orders of magnitude greater than 
$q_e$ in the layers with the optical depth $\sim 1$. In the region 
with a smaller optical depth, electron transport becomes a 
bit more efficient but still much less important than 
the radiative one. Therefore, $q_e$ plays no essential role 
in thermal balance and the first term on the r.h.s. of Eq.~(6) 
can be neglected. This is a qualitative difference to the 
thermomagnetic instability in laser-produced plasma 
where the electron thermal conductivity is the dominating 
mechanism of heat transport (see, e.g., Tidman \& Shanny 1974, 
Bissell 2015, Bissell et al. 2012, Brownell 1979, and Hirao \& 
Ogasawara 1981). As far as the contribution 
of small terms proportional to the electron thermal conductivity 
is concerned, it is easy to estimate that they give a correction 
to the dispersion equation of the order of a ratio (electron 
thermal conductivity)/(radiative thermal conductivity) that is 
extremely small in stellar condition.

The momentum and continuity equations read
\begin{eqnarray}
\frac{d \vec{v}}{dt}= - \frac{\nabla p}{\rho} + \vec{g}
+ \frac{1}{4 \pi} (\nabla \times \vec{B}) \times \vec{B}, \\
\frac{\partial \rho}{\partial t} + \nabla \cdot ( \rho \vec{v})
=0,
\end{eqnarray}
where $\vec{g}$ is gravity. In the case $\vec{q}_e \neq 0$, 
linearization of Eq.(6) and Eqs.(7)-(8) leads to the set of 
equations containing disturbances of $T_1$, $\rho_1$, $p_1$, 
$\vec{B}_1$, and $\vec{v}_1$. However, this set does not 
contain $\vec{B}_1$ if $q_e$ is neglected. Indeed, if $q_e$ is 
negligible, the only term in Eqs.~(6) and (7)-(8) that depends 
on $\vec{B}_1$ is the Lorentz force in 
the momentum equation (7). Linearizing the Lorentz force, 
we have $[(\nabla \times \vec{B}) \times \vec{B}]_1 = (\nabla 
\times \vec{B}_1) \times \vec{B} + (\nabla \times \vec{B}) 
\times \vec{B}_1 =0$ since $\vec{B} = 0$ in the basic state. 
Therefore, Eqs.(6), (7) and (8) can be decoupled from Eq.(5) 
and these equations have their own set of eigenmodes and 
eigenvalues. On the other hand, the particular type of eigenmodes 
caused by the thermomagnetic effects can be described by Eq.(5) 
alone if one supposes that $T_1$, $\rho_1$, $p_1$, and $\vec{v}_1$ 
are vanishing. Then, only disturbances of $\vec{B}_1$ are 
non-vanishing in this particular mode that can be called 
thermomagnetic. In the case $\vec{q}_e \approx 0$, the equation 
governing this mode reads 
\begin{eqnarray}
\frac{\partial \vec{B}_1}{\partial t} = \!-\! 0.81 \frac{k_B}{m_e} 
\nabla \! \times \! (\tau_e \nabla T \!\times\! \vec{B}_1)
- \frac{c^2}{4 \pi} \nabla \!\times\! \left( \frac{1}{\sigma}
\nabla \!\times \! \vec{B}_1 \right).
\end{eqnarray}   
The thermomagnetic modes exist if the temperature
is non-uniform. Comparing the first and 
second terms on the r.h.s. of Eq.(10) and assuming that the 
lendth-scales of perturbations and unperturbed quantities are
of the same order, we obtain that thermomagnetic effects 
yield a stronger influence than ohmic dissipation if 
\begin{equation}
\varepsilon \equiv \frac{c_e^2}{c^2} \; \omega_p^2 \tau_e^2 \gg 1,
\end{equation}
where $c_e = \sqrt{k_B T/m_e}$ is the thermal velocity of electrons
and $\omega_p = \sqrt{4 \pi e^2 n_e/m_e}$ the plasma frequency.
Then, Eq.~(11) yields
\begin{equation}
\varepsilon \approx 36 T_4^4 / n_{13} \Lambda^2  \gg 1,
\end{equation}
where $n_{13} = n/ 10^{13}$cm$^{-3}$ and $T_4 = T/10^4 K$. 
If the wavelength of perturbations is much shorter than the 
unperturbed length-scale, then the Nernst effect is
especially effective in the limit of long-wavelength perturbations
when dissipation is minimized (see Hirao \&
Ogasawara (1981) and Bissell (2015)).    

We consider Eq.(9) in the case $\varepsilon \gg 1$ 
when the thermomagnetic effects are important.
Note that thermomagnetic modes have a simple form (9) only 
in a linear approximation when one does not take into account 
the Hall parameter and Lorenz force.

\section{Thermomagnetic instability} 

It is seen from Eq.(14) that thermomagnetic effects can influence 
only the component of $\vec{B}_1$ perpendicular to $\nabla T$. It 
is convenient to choose the $y$-axis parallel to $\vec{B}_1$. If 
the basic state is quasi-stationary, $\vec{B}_1$ depends
on $t$ and $x$ as $\propto \exp(\gamma t - i k_x x)$ where $\gamma$ 
is the growth rate and $k_x$ is the wavevector 
in the $x$-direction. The dependence on $z$ should be determined 
from Eq.(14). Under these assumptions, the equation governing 
the magnetic field reads
\begin{equation}
\eta_m B_{1y}^{''} + A B_{1y}^{'} + D B_{1y} = 0, 
\end{equation} 
where $\eta_m = c^2 / 4 \pi \sigma$ is the magnetic diffusivity
and
\begin{eqnarray}
A = 0.81 \frac{k_B}{m_e} \tau_e \frac{d T}{d z} \!-\!
\eta_m \frac{d \ln \sigma}{d z} , \;\;\;\;\;\;\;\;\;\;
\nonumber \\ 
D = D_0 \!-\! \eta_m k^2_x \!-\! \gamma , \;\;\; D_0 =
0.81 \frac{k_B}{m_e} \frac{d}{dz} \left(\tau_e \frac{d T}{d z} \right);  
\end{eqnarray}
the prime denotes $d /d z$. The ratio of the first and second 
terms on the r.h.s. of the expression for $A$ is of the order 
of $\varepsilon$ and, therefore, the second term ($\propto 
(d \ln \sigma/ dz)$) can be neglected in the region where 
thermomagnetic effects play a dominant role. 

To illustrate the main qualitative features of thermomagnetic
modes, we consider the behaviour of disturbances with a 
wavelength in the $z$-direction that is much shorter than the 
characteristic lengthscales of unperturbed quantities. Then the 
solution of Eq.(13) is $\propto \exp[i \int q(z) dz]$ and we 
have
\begin{equation}
B_{1y} = F_1 \exp \left[ i \int q_1(z) dz \right]  + 
F_2 \exp \left[ i \int q_2(z) dz \right],
\end{equation} 
where $F_1$ and $F_2$ are constants that must be determined 
from the boudary conditions. The equation for $q_{1,2}$ reads
\begin{equation}
q^2 - \frac{iA}{\eta_m} q  - \frac{D}{\eta_m} -i q' = 0. 
\end{equation}
If the wavelength is small, the last term on the 
l.h.s. is also small compared to the first one and it can be considered 
as a perturbation. Therefore, Eq.(16) can be solved by making use
of the standard perturbation procedure. If we represent the 
quantity $q$ as a sum of subsequent perturbation terms, $q = q^{(0)}
+ q^{(1)} + ...$, then the first two terms in this expansion satisfy
the following equation
\begin{equation}
q^{(0)\; 2} - \frac{iA}{\eta_m} q^{(0)}  - \frac{D}{\eta_m} = 0, \;\;\;
\;\; q^{(1)} = \frac{i q'^{(0)}}{2 q^{(0)} -i A/ \eta_m}. 
\end{equation} 
These two terms are sufficient for our consideration. It follows from
Eq.(17) that there are two thermomagnetic modes and  $q^{(0)}$ for 
these modes is given by
\begin{equation}
q^{(0)}_{1,2} = \frac{iA}{2 \eta_m} \pm \sqrt{ \frac{D}{\eta_m} -
\frac{A^2}{4 \eta^{2}_m}}.
\end{equation}
Estimating different terms in $q_1$ and $q_2$, one has $A \sim c_e^2 
\tau_e / L$ and $D \sim c_e^2 \tau_e / L^2$ where $L$ is the vertical 
lengthscale. Since $(D/\eta_m)/(A^2/4 \eta_m^2) \ll 1$, we have
\begin{equation}
\sqrt{\frac{D}{\eta_m} - \frac{A^2}{4 \eta^{2}_m}} \approx
\frac{iA}{2 \eta_m} \left( 1 - \frac{2 \eta_m D}{A^2} \right).
\end{equation}
Substituting this expression into Eq.~(18), we obtain
\begin{equation}
q_1^{(0)} \approx \frac{iA}{\eta_m} -\frac{iD}{A}, \;\;\;
q_2^{(0)} \approx \frac{iD}{A}.
\end{equation}

As it was noted, $\gamma$, $F_1$, and $F_2$ should be determined 
from the boundary conditions. Generally, the results depend on 
boundary conditions but this 
dependence is not very essential for short wavelength modes. 
Therefore, we consider the simplest case assuming that $B_{1y}$ 
is vanishing at the bottom (($B_{1y}=0$ at $z=0$) and 
the electric current is zero at the top ($dB_{1y}/dz=0$ at $z=a$). This 
choice seems to be plausible because the thermal generation cannot 
operate in deep layers where the rates of thermomagnetic processes 
becomes very low. Also, the considered mechanism generates only the 
toroidal magnetic field $B_{1y}$ and, hence, the electric current should 
vanish at the stellar surface. From the condition $B_{1y} = 0$ at 
$z=0$, we obtain $F_1 = -F_2$, and then 
\begin{equation}
B_{1y} = F_1 \left\{ \exp \left[ i \int_0^z q_1(z) dz \right]  - 
\exp \left[ i \int_0^z q_2(z) dz \right] \right\}.
\end{equation} 
The second boundary condition ($dB_{1y}/dz=0$ at $z=a$) yields
\begin{equation}
q_1(a)  - q_2(a) \exp \left[ i \int_0^a [q_2(z)-q_1(z)] dz \right] =0,
\end{equation} 
By making use of Eq.~(20), we obtain
\begin{equation}
q_1(a)  - q_2(a) \exp \left[ \int_0^a (A/\eta_m) dz \right] =0.
\end{equation} 
The solution of this equation is crucially dependent on the 
sign of $A$ that is determined by $dT/dz$. If $dT/dz <0$ and the 
temperature decreases outward then the 
exponential term on the l.h.s. of Eq.(23) is small since 
$|\int_0^a (A/\eta_m) dz| \sim |aA/\eta_m| \sim (a/L) \varepsilon$ 
is large and negative in the region where $\varepsilon \gg 1$ and 
termomagnetic effects are important. Therefore, one can neglect 
the exponential term in Eq.(23) and the dispertion relation reads 
$q_1(a) \approx 0$ in this case. Then, using Eq.~(20), one has 
\begin{equation}
 \frac{A}{\eta_m} -\frac{D}{A} \approx 0
\end{equation}
 or, using Eq.(18),
\begin{equation}
\gamma \approx - \frac{A^2}{\eta_m} + D_0.
\end{equation} 
Since $A^2/\eta_m \gg D_0$ (see Eq.~(19) and the discussion 
above it), we obtain $\gamma \approx - A^2/\eta_m < 0$ and, hence, 
generation of the magnetic field is impossible if 
$dT/dz < 0$.

 In the region with the inverse temperature gradient, $dT/dz > 0$, 
the situation differs drastically. In this case, $A$ is positive
and the second term on the l.h.s. of Eq.~(23) gives a dominating
contribution since the exponent is proportional to 
$\int_0^a (A/\eta_m) dz \sim aA/\eta_m \sim (a/L) \varepsilon \gg 1$. 
Therefore, the first term can be neglected and the dispersion 
relation reads $q_2(a) \approx 0$ or $D(a)=0$. Then,
\begin{equation}
\gamma \approx  D_0(a) \approx 0.81 \frac{k_B}{m_e} \frac{d}{dz} 
\left(\tau_e \frac{d T}{d z} \right) - \eta_m k_x^2.
\end{equation}
Taking into account that $\tau_e \propto T^{3/2}/n$ (Spitzer 1998), 
we obtain
\begin{equation}
\gamma \sim 0.81 \frac{k_B}{m_e} \tau_e \left[ \frac{3}{2T} \!
\left( \frac{dT}{dz} \right)^2  \!-\! \frac{d \ln \rho}{dz} \frac{dT}{dz}
       \!+\! \frac{d^2 T}{dz^2} \right] \!-\!  \eta_m k_x^2 .
\end{equation}
Generally, $\gamma$  can be positive or negative depending on the
temperature and density profiles. It appears that $\gamma$ is 
positive in some layers with the inverse temperature 
gradient and the instability occurs in these layers. Such layers 
are formed in massive stars because of their high luminosity and 
departures from the local thermodynamic equilibrium in the upper 
atmosphere. This behaviour of $T$ is well known from the atmospheric 
models developed first by Auer \& Mihalas (1969a,b) and confirmed 
later by many authors (see, e.g., Gabler et al. 1989 and Martins 
2004 for review). Because of a departure from the local 
thermodynamic equilibrium, the temperature profile in the atmosphere 
of massive stars has a bump-like structure in the region where the 
optical depth is less than $1$ (see, e.g., Fig. 2.1 after Martins
2004). The bump in a temperature profile is typically located at 
the optical depth $\sim 0.01-0.001$ and its height depends on the 
surface temperature. In our model, we can assume that the bottom of 
a generating layer ($z=0$ in our notations) corresponds to the 
depth where the temperature gradient changes the sign. Then, it is 
easy to check that $\gamma > 0$, at least in a fraction of the bump 
region. Indeed, the first term in the brackets on the r.h.s. of 
Eq.(27) is always positive. The second term is also positive in 
layers with the inverse temperature profile ($dT/dz >0$). The third 
term, $d^2T/dz^2$, should be positive in a some region with $dT/dz 
> 0$ because of a bump-like shape of the temperature profile. 
Hence, there always exist the region where $\gamma >0$ and the 
instability occurs. The 
characteristic lengthscale of this region is comparable to the 
thickness of a layer with the inverse temperature gradient.

\section{Discussion}

This paper considers the thermal generation of the magnetic field 
in surface layers of massive stars. Such generation is possible 
in the upper atmosphere of hot stars where departures from the 
local thermodynamic equilibrium form a region with the inverse 
temperature gradient. Such a behaviour of $T$ is well known from 
atmospheric modelling and is rather general in massive stars (see, 
e.g., Auer \& Mihalas 1969a,b, Martins 2004). An effectivity of the 
thermal generation is determined by value of the inverse temperature 
gragient and thickness of this layer. The 
inverse temperature gradient exists typically in layers with the 
optical depth $\tau < 0.01-0.001$. 

We have considered only generation of small-scale magnetic 
fields with the horizontal wavelengths $\lambda = 2 \pi / k_x$  
shorter than the lengthscale of unperturbed quantities, $L$. Note, 
however, that the considered mechanism can generate magnetic fields 
with the lengthscale comparable to $L$ as well. Certainly, unstable
perturbations do not have a wavelike shape in this case. For instance, 
generation of such fields has been studied by Urpin et al. (1986) 
in neutron stars. It follows from Eq.(27) that, in a short-wavelength 
approximation ($L \gg \lambda$), $\gamma$ reaches its maximum if $k 
\rightarrow 0$ (or $\lambda \rightarrow \infty$). However, Eq.(27) 
does not apply if $\lambda \geq L$ and the maximum lengthscale of 
unstable perturbations is restrickted by the lenghtscale 
of unperturbed quantities, $L$. Numerical calculations by Urpin et al. (1986)
for $\lambda \geq L$ confirm this qualitative conclusion. Indeed, $\gamma 
\rightarrow 0$ if $\lambda \rightarrow \infty$ and the maximum value of 
the growth rate is reached at $\lambda \sim L$ and is given by Eq.(27) 
at $\lambda \sim L$.
 
The condition of instability for short-wavelenghts reads $\gamma > 0$ 
and, hence, the minimum unstable wavelength can be estimated as 
$\lambda_{min} \sim L \varepsilon^{-1/2}$. Therefore, the range of 
unstable wavelengths is $\lambda_{min} < \lambda < L$, and these fields 
can manifest themselves as magnetic spots on the surface. Such spots 
are in qualitative agreement with the lengthscale of observed magnetic 
spots. Indeed, if $T=3 \times 10^4$, then 
$\varepsilon^{-1/2} \sim 30$ in the atmosphere where $n_{13} =1$.
Since, $L \sim 10^9$cm in massive stars, generated magnetic spots
have a lengthscale $\simeq 3 \times 10^{10}$ cm.

By making use of Eq.(27), the growth rate of magnetic disturbances 
can be estimated  as    
\begin{equation}
\gamma \sim (k_B/m_e) \tau_e (3/2T) 
\left( dT/dz) \right)^2 \sim c_e (\lambda_e/L^2),
\end{equation}
where $\lambda_e =c_e \tau_e$ is the mean free-parth of electrons.
The standard expression for the electron relaxation time (Spitzer 
1998) yields the following estimate for the growth time 
$t_B = 1/ \gamma$ of disturbances
\begin{equation}
t_B \sim 10^3 \; n_{13} \; L_9^2 \; \Lambda T_{4}^{-5/2} {\rm yrs},
\end{equation}   
where $L_9 = L/10^9$cm. For typical parameters ($n_{13}=L_9=1$, 
$\Lambda=4$, $T_4 = 3$), we obtain 
$t_B \sim 10^4-10^5$ yrs. The proposed mechanism cannot
generate the magnetic field in very young stars (with the age 
$< 10^4$ yrs). The generation is possible, however, for older 
stars. The main-sequence lifetime of massive stars, $\tau_{ms}$, is 
relatively short but, nevertheless, the timescale of the magnetic 
field generation can be shorter. The lifetime of massive stars is 
\begin{equation}
\tau_{ms} \approx 10^{10} (M/M_{\odot})^{-2.5} {\rm yr},
\end{equation}  
where $M$ is the stellar mass (see, e.g., Bhattacharya \& van den
Heuvel 1991, Urpin et al. 1998). The lifetime of very massive stars 
(with $M > 100 M_{\odot}$) is obviously shorter than $t_B$ and, 
hence, the field cannot be generated in very massive stars. As far 
as less massive stars are concerned, generation by the thermal 
mechanism is possible for such stars.  

Thermomagnetic waves have a simple form with vanishing disturbances 
of all quantities except $B_{1y}$ only in a linear regime and, hence, 
the thermomagnetic instability leads to an exponential growth of the 
magnetic field only in this regime. Likely, the magnetic disturbances 
play a crucial role in settling down a saturation regime. The thermal  
generation is determined by the Nernst effect and caused by the term 
proportional to $\beta_{\wedge}$ in induction equation (3). This 
generating effect is suppressed if the magnetic field becomes so 
strong that the Hall parameters $x_e \omega_B \tau_e \sim 1$. Likely, 
this condition yields an estimate of the saturation regime of the 
thermomagnetic instability. Note that laboratory experiments and 
numerical modelling (see, e.g., Tidman \& Shanny (1974), Bol'shov, 
Dreizin, \& Dykhne (1974), and Andrushchenko \& Pavlenko (2004)) 
are in a qualitative agreement with this simple estimate of a 
saturation state. The condition $x_e \omega_B \tau_e \sim 1$ yields 
the following estimate for a saturation magnetic field
\begin{equation} 
B_{sat} = \frac{m_e c}{e \tau_e} \sim 10^2 \; \frac{n_{13} 
\Lambda}{T_4^{3/2}} \;\;\;{\rm G}.
\end{equation}
In stellar interiors, the ratio $n/T^{3/2}$ usually increases
with the depth from the surface and, therefore, the saturation 
field is stronger in deep layers. Eq.~(31) yields the magnetic
field that is in a qualitative agreement with that observed
in massive stars.

{\it Acknowledgements.} The author thanks the Russian
Academy of Sciences for financial support under the 
programme OFN-15.

\end{document}